**Topical Review**

# Computational electron-phonon superconductivity: from theoretical physics to material science

Shiya Chen[1], Feng Zheng[2*], Zhen Zhang[3], Shunqing Wu[1],
Kai-Ming Ho[3], Vladimir Antropov[4], Yang Sun[1*]

[1]Department of Physics, Xiamen University, Xiamen, 361005, China
[2]School of Science, Jimei University, Xiamen, 361021, China
[3]Department of Physics and Astronomy, Iowa State University, Ames, IA, 50011, USA
[4]Ames National Laboratory, U.S. Department of Energy, Ames, IA, 50011, USA

## Abstract

The search for room-temperature superconductors is a major challenge in modern physics. The discovery of copper-oxide superconductors in 1986 brought hope but also revealed complex mechanisms that are difficult to analyze and compute. In contrast, the traditional electron-phonon coupling (EPC) mechanism facilitated the practical realization of superconductivity in metallic hydrogen. Since 2015, the discovery of new hydrogen compounds has shown that EPC can enable room-temperature superconductivity under high pressures, driving extensive research. Advances in computational capabilities, especially exascale computing, now allow for the exploration of millions of materials. This paper reviews newly predicted superconducting systems in 2023-2024, focusing on hydrides, boron-carbon systems, and compounds with nitrogen, carbon, and pure metals. Although many computationally predicted high-$T_c$ superconductors were not experimentally confirmed, some low-temperature superconductors were successfully synthesized. This paper provides a review of these developments and future research directions.

[*]Email: fzheng@jmu.edu.cn (F.Z.); yangsun@xmu.edu.cn (Y.S.)

## I. Introduction

Searching for room-temperature superconductivity (SC) is recognized as one of the most significant challenges in modern physics. The discovery of copper-oxygen superconductors in 1986 sparked hope that this long-sought dream might be realized [1]. However, understanding the physics behind these superconductors was proven to be complex, turning this area into a vast and challenging playground for theoretical physicists. Spin fluctuations (SF), believed to be major contributors to this type of SC, pose significant analytical and computational difficulties [2,3] due to their wide frequency range, which exceeds that of phonon excitations (<0.2 eV). Consequently, material science analysis of these superconductors is currently not feasible.

In contrast, the electron-phonon coupling (EPC) mechanism of SC, described by the Bardeen-Cooper-Schrieffer (BCS) theory [4], allows for a clearer separation of interactions into the local Coulomb pseudopotential, making SC calculations feasible. This theoretical framework facilitated the practical realization of high-pressure superconductivity in metallic hydrogen, as suggested by Abrikosov [5], De Gennes [6], and Ashcroft [7] in the 1960s. However, until around 2015, EPC was not considered the primary mechanism for achieving room-temperature SC. The discovery of superconducting hydrogen compounds under pressure demonstrated that EPC could indeed be a source of room temperature SC, leading to the development and widespread use of computational codes based on the theory that incorporates both electron and phonon spectra of real materials[8–14]. These computational advancements spurred extensive searches for EPC superconducting materials [15], leading to numerous confirmed discoveries such as $CaH_6$ [16], $YH_6$ [17], $YH_9$ [18], and $LaH_{10}$ [19,20]. A comprehensive review article [21] recently summarized the studies of high-pressure metal superhydrides superconductors up to mid-2023.

The exponential growth in computing power over the past few years, culminating in the advent of exascale computing systems capable of performing $10^{18}$ calculations per second [22,23], also enables unprecedentedly fast calculations. These systems, along with many-core architectures in modern CPUs offering high core counts and advanced parallel processing capabilities, are ideal for high-throughput tasks such as searching for SC in millions of materials. In addition, the development of quantum simulators [24,25] are also advancing the understanding of microscopic origin of superconductivity.

The development of efficient computational schemes for total energy and EPC SC calculations forms the basis for computationally searching for SC. The stability of compounds can now be reasonably evaluated using convex hull and phonon calculations. Crystal structure prediction (CSP) methods such as CALYPSO [26], USPEX [27], AIRSS [28], and high-throughput screening techniques [29–32], enabled researchers to systematically explore extensive configurational spaces to identify stable crystalline structures under pressures. The strength of EPC and $T_c$ can also be directly computed with well-optimized community codes, yielding qualitatively consistent results. Machine learning techniques have been implemented to further accelerate $T_c$ calculations. The workflow and related algorithms for predicting EPC SC are well summarized in a recent review by Pickett [33].

It is now evident that computational material science can reliably predict EPC SC, provided the electronic structure is adequately described. As a result, there is an intense search for high $T_c$ EPC SC under ambient pressures, with a few systems predicted to have $T_c$ values close to or exceeding that of nitrogen in the past year. This review summarizes the newly predicted superconducting systems from 2023 to 2024, focusing on their chemical compositions and structural motifs.

## II. Predicted superconductors in 2023-2024

We identified over a hundred papers on computational predictions of superconducting systems from 2023 to 2024 using the Web of Science database. Our analysis of the chemical compositions revealed that most predicted systems are hydrides, B-C systems, or compounds containing nitrogen, carbon, or pure metal elements. This section classifies and analyzes the motifs of these compounds, grouping them within each category.

### 2.1 Predictions of hydride superconductors

Figure 1 presents the $T_c$ of newly predicted hydride superconductors at various pressures. It reveals that hydride superconductors in the lower-pressure region have lower hydrogen content and lower $T_c$ compared to those in the higher-pressure region. It was suggested that, at ambient pressure, hydrogen atoms typically exist as $H_2$ molecules, lacking the metallic properties and strong ionic behavior required for superconductivity [34,35]. In contrast, at high pressure, systems with higher $T_c$ values undergo the dissociation of hydrogen molecules into hydrogen ions. This pressure-induced change alters the material's electronic band structure, resulting in free electrons and strong electron-ion interactions necessary for superconductivity [33]. Systems with $T_c$ values exceeding 200 K typically have a hydrogen content higher than 0.8.

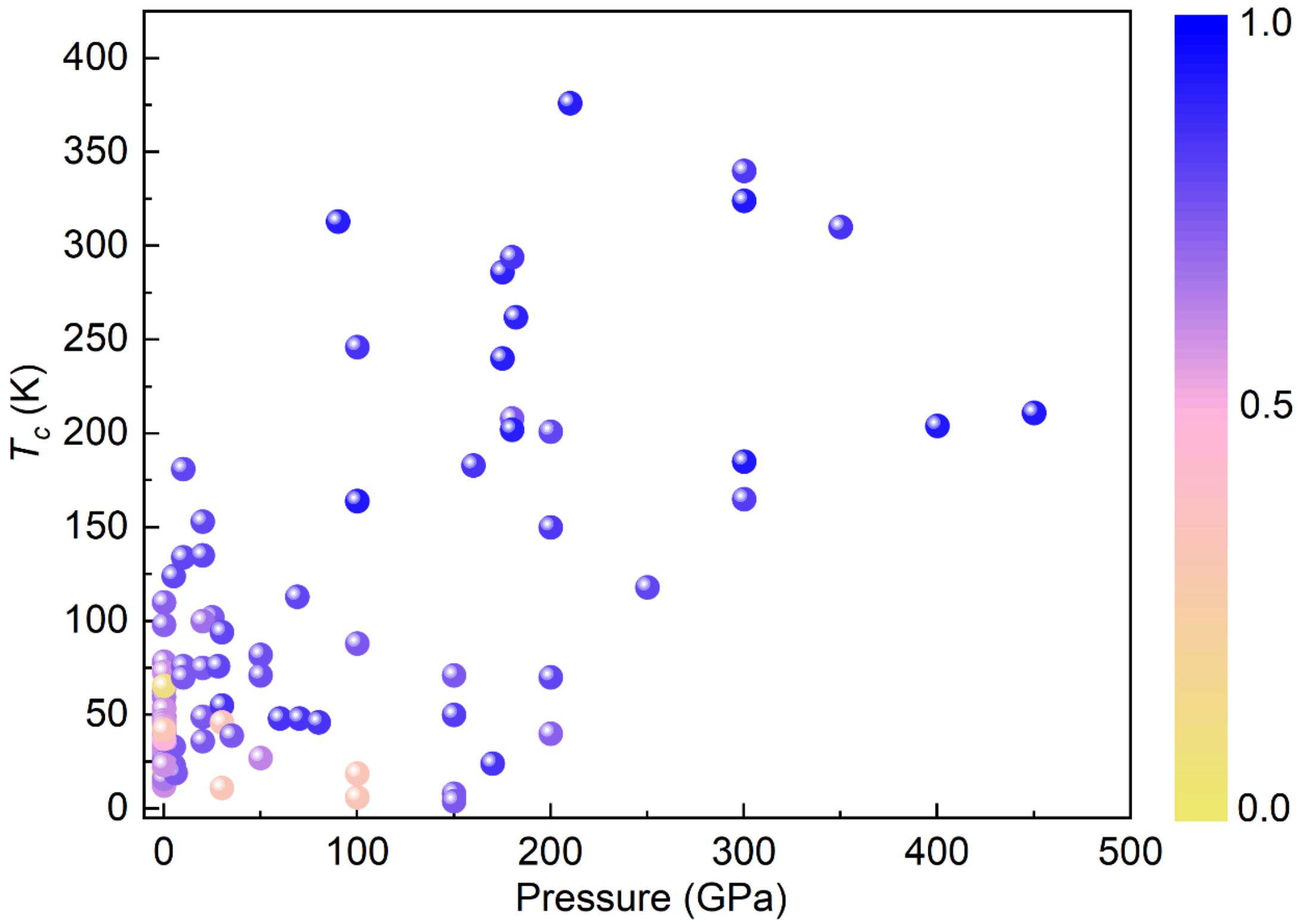

Fig. 1 $T_c$ and corresponding pressure of newly reported hydride systems in 2023-2024 [36–60]. The color in the point indicates the H percentage in each compound.

As shown in Fig. 2, the hydride superconductors recently predicted under ambient pressure can be classified into three main categories, including the $X_2YH_6$ system (where X represents rare earth or alkali metals, and Y represents transition or inert metals), the perovskite system (represented by $KInH_3$), and the $Y(BH_4)_2$ system. These three categories share a common structural characteristic: metal and hydrogen atoms form an octahedral packing motif.

**2.1.1 $X_2YH_6$**

The $X_2YH_6$ system exhibits a face-centered cubic lattice structure, with hydrogen atoms surrounding transition metal atoms to form isolated octahedra. A representative compound, $Mg_2IrH_6$, was discovered by Dolui et al. through high-throughput searches of various highly symmetric ternary hydrides [61]. They proposed a feasible synthesis route for $Mg_2IrH_6$ by first synthesizing the insulating phase $Mg_2IrH_7$ under high pressure (15 GPa) and then dehydrogenating it. Sana et al. predicted that $Mg_2PtH_6$, belonging to the same structural family, could achieve a $T_c$ exceeding 100 K under ambient pressure with electron doping [62]. Further, Zheng et al. [63] identified more $X_2YH_6$ systems, suggesting that these systems exhibit sharp Van Hove singularities at their Fermi levels, contributing to their high $T_c$. Wang et al. showed that replacing Mg with Ca resulted in the disappearance of superconductivity in $Mg_2IrH_6$, suggesting that the rigid antibonding in the $IrH_6$ polyhedron may be key to the high $T_c$ in $X_2YH_6$ systems [64].

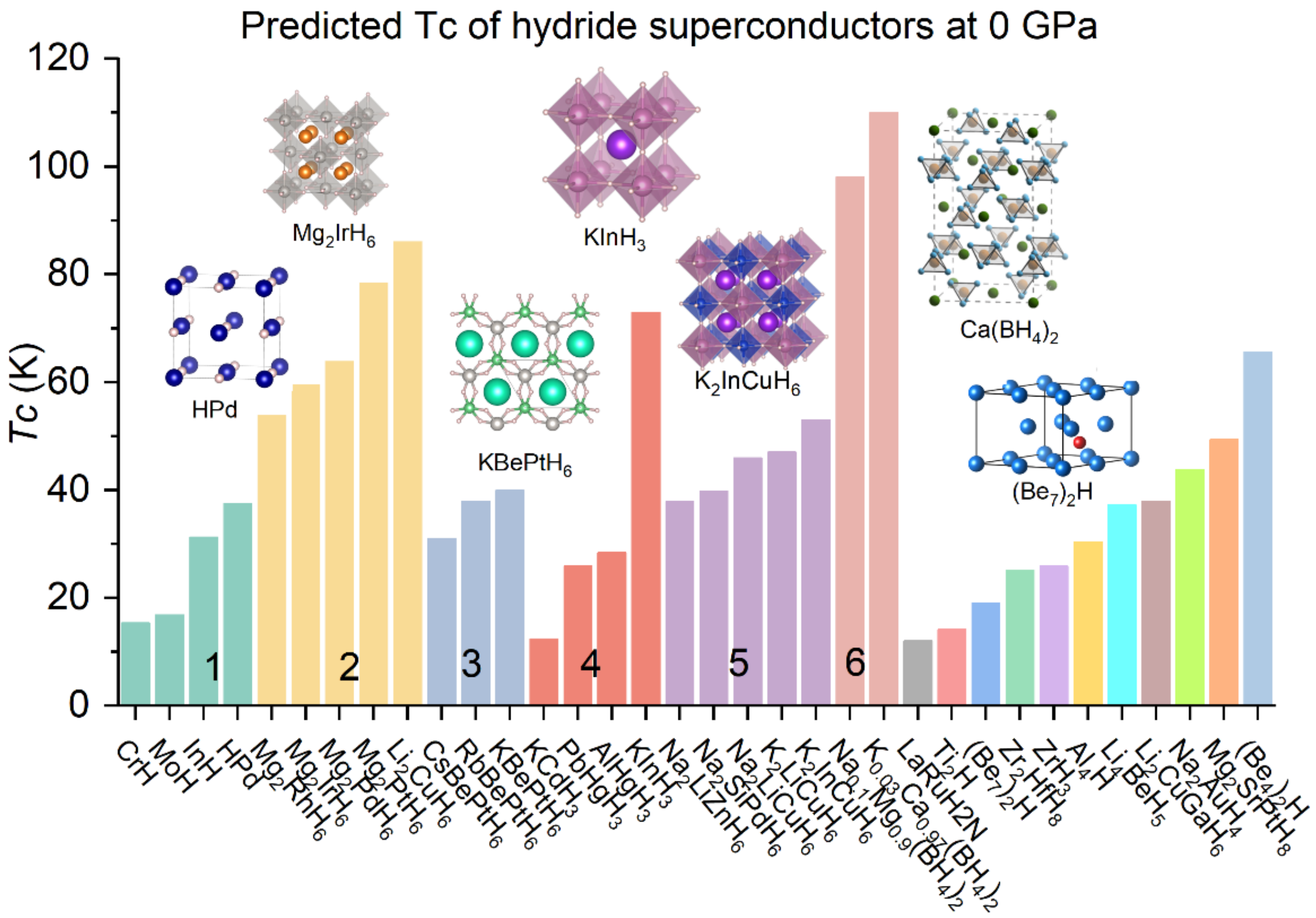

Fig. 2 Predicted ambient hydride superconductors in 2023-2024. Different colors in the columns indicate various structural motifs. Data for type 1 come from Ref.[36,37], type 2, type 3, type 4, and type 5 from Ref.[37], and type 6 from Ref.[57,65]. The remaining colorful columns correspond to different types of structures, as reported in Ref.[36–38].

### 2.1.2 Perovskite Structure

The ideal perovskite structure, which exhibits cubic symmetry, has been linked to superconductivity since the discovery of the high-$T_c$ superconductor $La_{2-x}Ba_xCuO_4$ by Bednorz et al. in 1986 [1]. Despite the mechanism still not being fully understood, its $T_c$ far exceeded that of conventional superconductors known at the time, sparking significant exploration in the field.

Recent research focused on perovskite-type superconductors, including the discovery of the non-oxide perovskite superconductor, $MgCNi_3$, by He et al [66]. Subsequent computational studies confirmed it as a conventional superconductor dominated by strong EPC [67]. In 2023, Tian et al. predicted the superconducting ternary metal hydride perovskite system, $MgHCu_3$ [68], with a $T_c$ of 42 K. Figure 2 shows that many perovskite-type conventional superconductors with higher $T_c$ predicted in the past two years, including the metastable phase $KInH_3$ [37], which is predicted to have a $T_c$ of 73 K. In these systems, the main contribution to EPC comes from low-frequency phonons induced by heavy atoms, exhibiting diverse electronic band structures and Fermi surfaces [23]. Although several hydride perovskite systems were synthesized, the predicted superconductors were not synthesized [69–71]. Many double perovskite hydride systems were predicted to exhibit superconductivity, theoretically proposed as high-temperature ferromagnetic semiconductors [72], but no system of this type was experimentally synthesized.

#### 2.1.3 $X(BH_4)_2$

In Figure 2, the doped $Ca(BH_4)_2$ system, a molecular crystal where boron and hydrogen form $BH_4$ tetrahedra, is predicted to exhibit a $T_c$ exceeding 100 K under ambient pressure [65]. Before doping, $Ca(BH_4)_2$ is an insulator, and a doping concentration as low as 0.03 holes per formula unit is sufficient to induce a metallic state. The high $T_c$ is attributed to the strong EPC between B-H σ molecular orbitals and bond-stretching phonons. Earlier this year, An et al. predicted a closely related system, Na-doped $Mg(BH_4)_2$ [57], with a $T_c$ reaching 98 K at a doping concentration of 0.1 holes per formula unit, potentially increasing to 140 K with higher doping concentrations. This system shares a similar superconducting mechanism with K-doped $Ca(BH_4)_2$. Other metal borohydrides, such as $KB_2H_8$ [22] and $CsB_2H_8$ [74], were also predicted to exhibit superconductivity with $T_c$ values of approximately 140 K at 12 GPa and 100 K at 25 GPa, respectively. Due to the wide commercial availability of metal borohydrides, experimental verification of these systems might be conducted promptly.

Moreover, $(Be_4)_2H$ [38], comprising hexagonal layers of Be and B, was predicted to exhibit conventional superconductivity under ambient pressure with an estimated $T_c$ approaching 70 K. However, this system is thermodynamically unstable, posing significant challenges for experimental synthesis and characterization.

### 2.2 Predictions of B-C superconductors

Figure 3 shows that the newly predicted B-C system superconductors at ambient pressure are primarily represented by two categories. The first extensively studied predicted superconductor, $Li_{1/2}BC$ with a $T_c$ above 40 K [75], originates from LiBC through hole doping. Layered LiBC compounds, akin to graphite, long captivated researchers due to their composition of light elements, suggesting strong covalent bonds and distinctive phonon frequencies, which could potentially lead to high-temperature superconducting properties. Their crystal structure resembles the conventional superconductor $MgB_2$, exhibiting $P6_3/mmc$ symmetry [76,77]. Unlike $MgB_2$, LiBC is an insulator [77]. Therefore, methods such as hole doping and the introduction of other elements were explored to induce metallicity in LiBC [78–80]. Zheng et al. extensively investigated the structure and superconducting properties of the Li-B-C system at 100 GPa, discovering that even in many non-layered Li-B-C systems, superconductivity can also be observed [81].

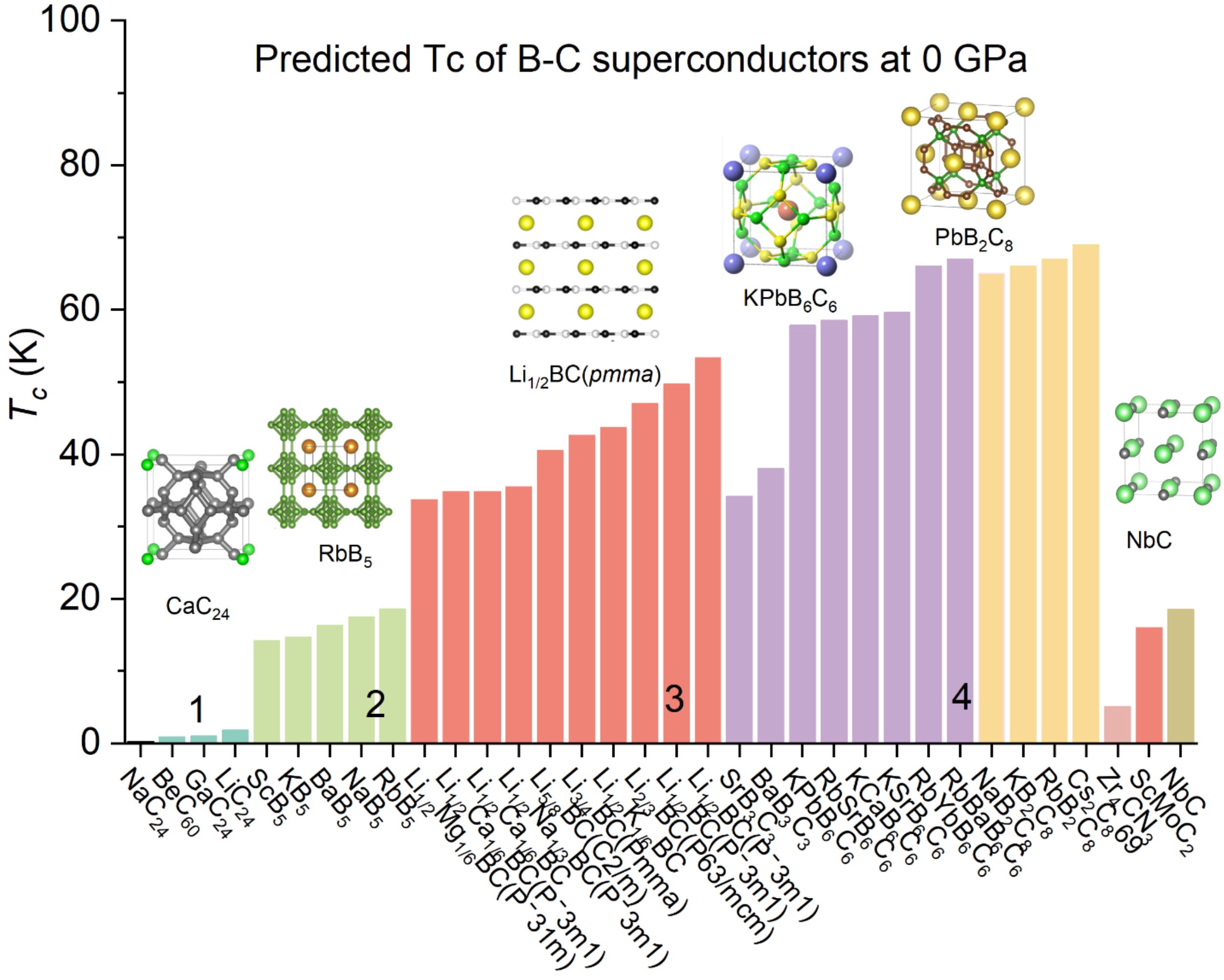

Fig. 3 The predicted ambient superconductors of B-C systems for 2023-2024. Different colors in the columns indicate various structural motifs. Data for type 1 come from Ref.[82], type 2 from Ref.[83], type 3 from Ref.[75], type 4 from Ref.[84,85], and type 5 from Ref.[86]. The remaining colorful columns correspond to different types of structures, as reported in Ref.[36,87].

Inspired by the synthesis of $SrB_3C_3$ under high pressure [88], which shares structural similarities with cage-like hydride superconductors currently under extensive study and is electronically dominated by a covalent B-C sublattice akin to the bonding and electronic properties observed in borides and carbides like $MgB_2$, extensive computational studies were conducted to explore its superconducting properties [89–91] and those of similar structures, such as $CaB_3C_3$ [89] ($T_c$ = 48 K at 0 GPa), $SrB_4C_2$ ($T_c$ = 19 K at 0 GPa)[92], $Rb_{0.4}Sr_{0.6}B_3C_3$ ($T_c$ = 83 K at 0 GPa) [93], $Rb_{0.5}S_{r0.5}B_3C_3$ ($T_c$ = 75K at 0GPa) [94] and $BaB_3C_3$ ($T_c$ = 43 K at 0 GPa) [90]. Over the past two years, research on this type of structure continued, exemplified by $KPbB_6C_6$ [84] ($T_c$ = 58 K at ambient pressure). High-throughput screening of all compounds with *Pm-3* symmetry, including $XB_3C_3$ and $XYB_6C_6$, revealed relatively high $T_c$ values. It is proposed that the $T_c$ of these compounds can be controlled by adjusting the average valence states of the X and Y metal elements. Duan et al. predicted that $RbBaB_6C_6$ and $RbYbB_6C_6$ could exhibit $T_c$ values of 67 K and 66 K, respectively, at ambient pressure [85]. The matching of the ionic radius of alkaline earth metal elements with the bond length of the cage-like B-C framework is identified as key for the kinetic stability of such compounds under ambient pressure, presenting a new avenue for identifying high-temperature superconductivity in covalent materials under environmentally stable conditions. Moreover, the $XB_2C_8$ (X = Na, K, Rb, Cs) [86] boron-carbon clathrates, exemplified by $CsB_2C_8$ (with a predicted $T_c$ of 69 K at 0 GPa), have also been identified as promising high-temperature superconductors. This discovery further expands the family of boron-carbon clathrate superconductors.

Moreover, Tomassetti et al. predicted a series of $Mg_xB_2C_2$ and $Na_xBC$ superconductors with $T_c$ ranging from 16 K to 29 K by performing hole doping through thermal deintercalation on $MgB_2C_2$ and NaBC ternary compounds [95]. Apart from these systems, the predicted $T_c$ of other newly proposed B-C system superconductors does not exceed 20 K. Among them, in the tetragonal crystal structure of pentaborides ($XB_5$) [83] (X = Na, K, Rb, Ca, Sr, Ba, Sc, and Y), $NaB_5$ exhibits the highest $T_c$ of 17 K, whereas the $T_c$ of sodalite-like structured ($XC_{24}$) [82] (X = Li, Na, K, Be, Mg, Ca, Al, Ga, Ge) is below 2 K. These structures all have clathrate-type motifs. Additionally, Chen et al. performed high-throughput screening on previously synthesized boride compounds and identified $TaMo_2B_2$ with a $T_c$ of 12 K [30]. At 40 GPa, Yang et al. predicted that $Sr_2B$ in the $F\bar{4}3m$ structure exhibits a superconducting transition temperature of 105 K [96].

## 2.3 Other predicted superconductors at ambient pressure

As shown in Fig. 4, there were a few computational predictions regarding other superconductors, with most systems derived from the work of Cerqueira et al. [36], who used machine-learning techniques to filter databases based on specific criteria. Among these, $LiMoN_2$ was predicted to exhibit a $T_c$ of 46 K [36] at ambient pressure, making it the system with the highest predicted $T_c$ at ambient pressure besides hydrides and B-C compounds. This system was experimentally synthesized in 1992 [97], but due to imperfect sample quality and structure, $T_c$ was not observed experimentally. Apart from hydrides and B-C systems, other compounds generally exhibited lower $T_c$ values. It was also observed that most predicted superconductors, apart from hydrides and borides, contained light elements such as C, N, O, and V. This observation aligns with BCS theory [4], where the presence of light elements in materials tended to result in higher Debye temperatures, potentially leading to higher $T_c$ superconductivity.

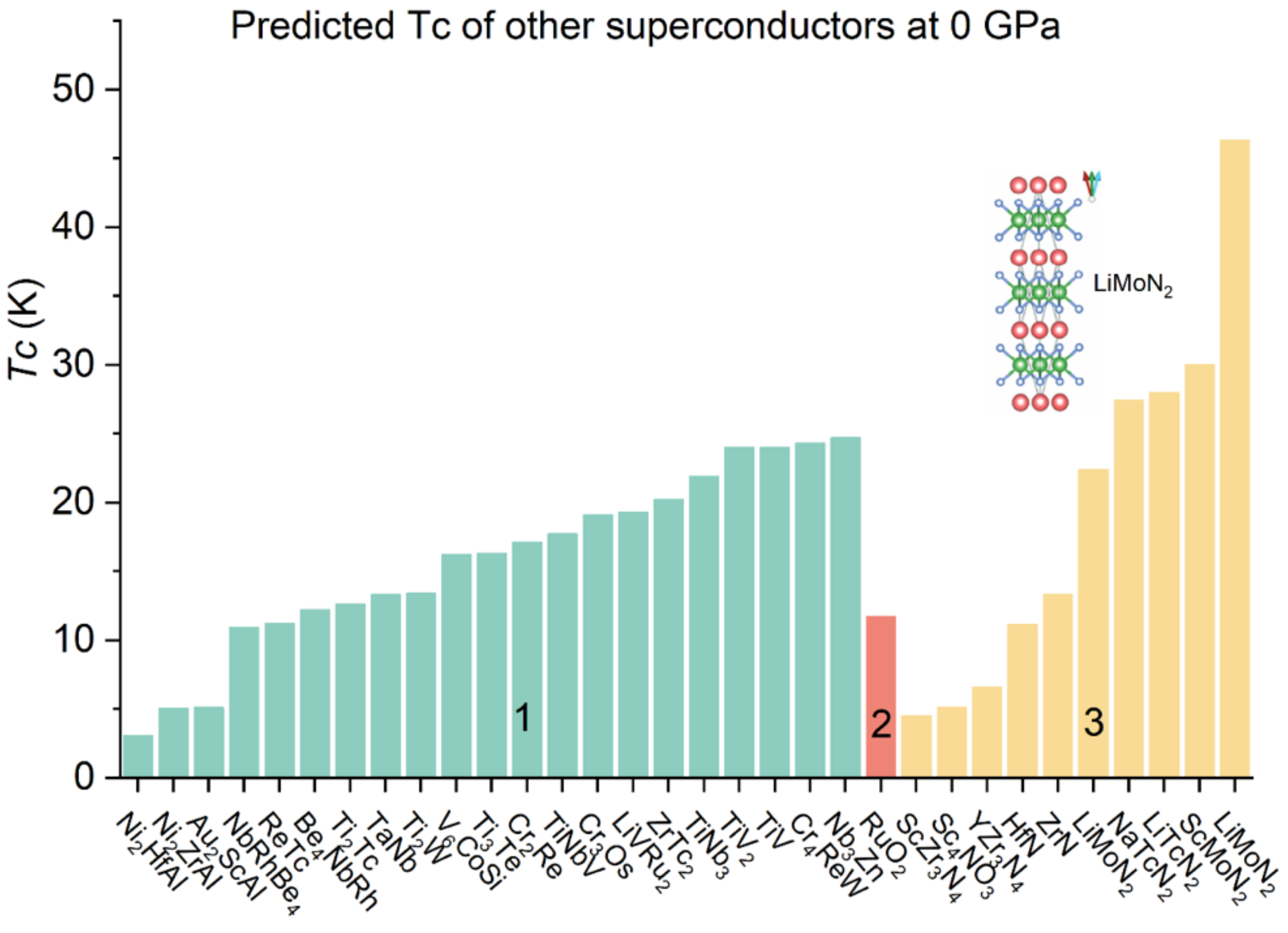

Fig. 4 The predicted ambient superconductors of systems excluding hydrides and B-C systems for 2023-2024 recorded from literature. Different colors in the columns indicate different chemical compositions. Type 1 represents intermetallics [36,87,98]. Type 2 represents oxides [36]. Type 3 represents nitrides [36,87].

## 2.4 Superconducting parameters

The computational calculations provide detailed data to analyze key parameters contributing to the $T_c$. We selected a few compounds from the $X_2MH_6$ [63], $XB_3C_3$, and $XYB_6C_6$ families [84] for analysis. Figure 5(a) and (b) summarizes the superconducting parameters, including the EPC constant (λ) and logarithmic average frequency ($\omega_{log}$) for selected $X_2MH_6$, $XB_3C_3$, and $XYB_6C_6$ compounds, arranged in order of decreasing $T_c$. The EPC contributions are divided into cations and H or B/C components. It is evident that cations significantly contribute to λ. In several compounds, the cation contribution to the EPC constant exceeds 30%, and in some cases, approaches ~50%, such as in $Al_2MnH_6$ and $KPbB_6C_6$.

We further analyze the correlation between $T_c$ and these superconducting parameters. Figure 5(c) suggests that λ has a very strong positive correlation with $T_c$ in both families, as suggested by the Pearson correlation coefficient $r$ of 0.95. In contrast, $\omega_{log}$ shows almost no correlation ($r$~0) with $T_c$ in $X_2MH_6$ system and a negative correlation with $T_c$ for $XB_3C_3$ and $XYB_6C_6$ compounds, as depicted in Fig. 5(d). To explain the different correlations, we separate the EPC contribution between metal and light elements (H or B/C) for these compounds. In the $XB_3C_3$ and $XYB_6C_6$ families, the B-C forms a rigid bipartite sodalite clathrate structure [84] and provides almost constant contributions to λ, as suggested by the green bars in Fig. 5(b). Thus, if the cation has a larger contribution to EPC, the total λ and $T_c$ should be larger. This is demonstrated by the strong positive correlation between $\lambda_M/\lambda_{BC}$ and $T_c$ shown in Fig. 5(e). Because the cation's phonon frequencies are much lower than those of B/C, and $\omega_{log}$ represents the frequency where strong EPC occurs, a larger EPC contribution from cation leads to a smaller $\omega_{log}$. Due to this mechanism, $XB_3C_3$ and $XYB_6C_6$ systems exhibit a strong negative correlation between $\omega_{log}$ and $T_c$, as shown in Fig. 5(d). However, in the case of $X_2MH_6$ system, the contributions to λ from H atoms vary among different compounds, as shown by green bars in Fig. 5(a). Because the phonon spectra of $X_2MH_6$ can significantly change by different X and M cations, particularly the phonon bands of H [37,62,63], it results in a weak correlation between the $\lambda_M/\lambda_H$ and $T_c$, as shown in Fig. 5(e). The mixed relationship among chemical elements, phonon spectra, and electron-phonon interactions weakens the correlation between $\omega_{log}$ and $T_c$ in $X_2MH_6$ system, as shown in Fig. 5(d). The factors that affect $T_c$ in $X_2MH_6$ system are quite subtle. For instance, we found isoelectronic substitution can significantly decrease $T_c$, when Ir was substituted by Co ($Mg_2CoH_6$ with $T_c$ = 36 K), or Mg was substituted by Sr ($Sr_2IrH_6$ with $T_c$ = 0 K). The mechanism for such effect is still under investigation for this novel system[64] These analyses demonstrate that the correlation between superconducting parameters and $T_c$ depends on the specific system. Finding a determining parameter to describe high-$T_c$ compounds remains a nontrivial task.

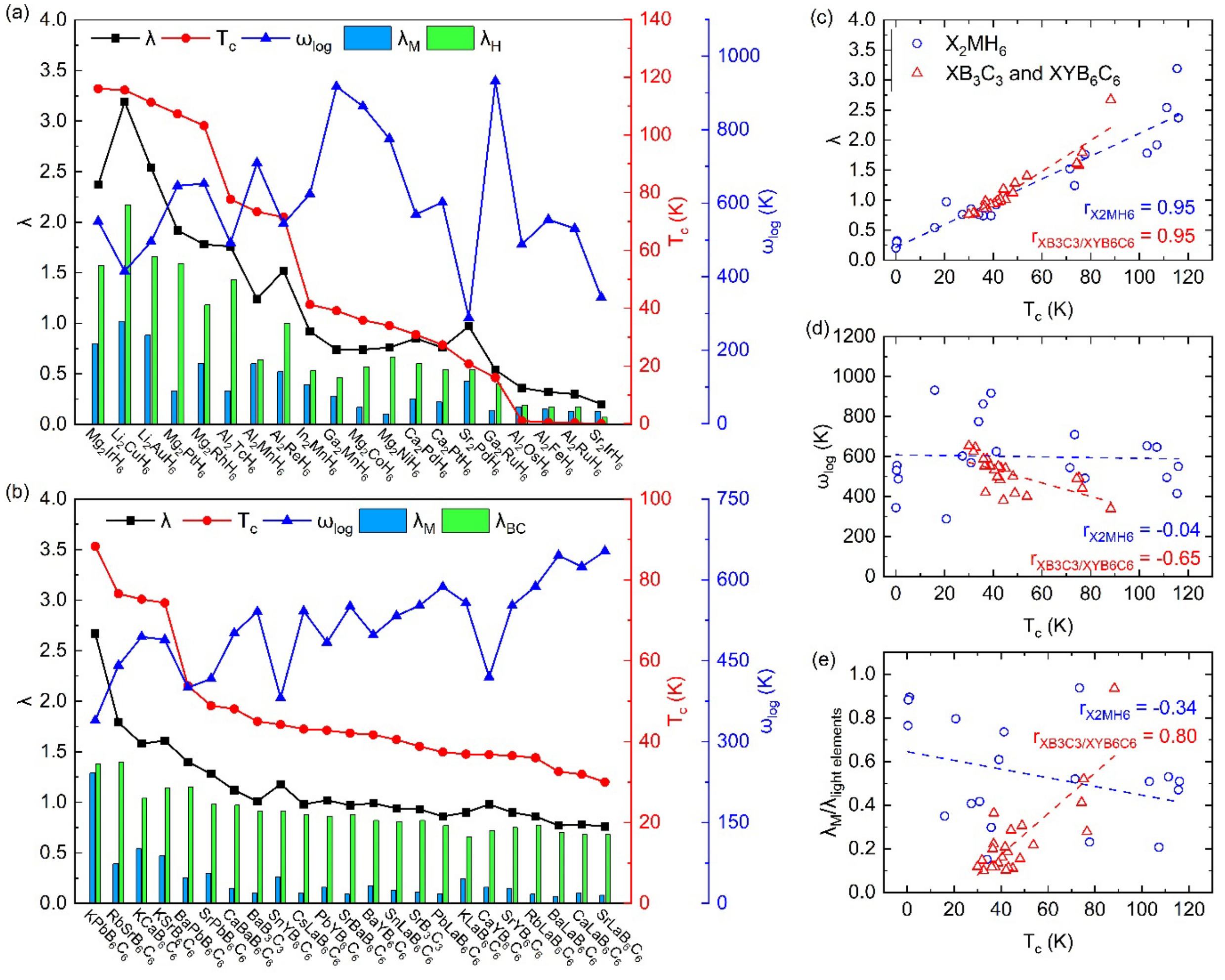

Fig. 5 Calculated superconducting parameters for (a) selected $X_2MH_6$ compounds and (b) $XB_3C_3$ and $XYB_6C_6$ compounds. The histograms denote the EPC parameter (λ) contributed from cations (M) and light elements (H or B/C). The data of $X_2MH_6$ compounds were obtained with Allen-Dynes correction using coupling and shape factors [63]. The data of the $XB_3C_3$ and $XYB_6C_6$ compounds were calculated by numerically solving the Eliashberg equations[84]. (c) The correlation between λ and $T_c$. (d) The correlation between logarithmic average frequency ($\omega_{log}$) and $T_c$. (e) Separated EPC contributions from metal and light elements ($\lambda_M/\lambda_H$ for $X_2MH_6$ family and $\lambda_M/\lambda_{BC}$ for $XB_3C_3$ and $XYB_6C_6$ family) and their correlations with $T_c$. $r$ is the Pearson correlation coefficient.

## III. Experimentally Synthesized Superconductors

Although many ambient-pressure superconductors are computationally predicted to have high $T_c$, experimental realization of these superconducting systems is rare. Figure 6 shows that in the past two years, there was no major breakthrough in synthesizing high-$T_c$ superconductors under ambient conditions. To date, $MgB_2$ with a $T_c$ of 39 K [99] remains the highest among conventional ambient-pressure superconductors. Within the low-pressure range, cuprates were the only unconventional superconductors with transition temperatures exceeding the liquid nitrogen temperature range before 2023 [21]. Several experimental advancements were reported at the high-pressure regime. Figure 6 highlights a region between 140 and 200 GPa and 150-260 K, which exhibits the relatively highest $T_c$ among the newly synthesized systems. In this range, the predicted $T_c$ values for $H_3S$ [100], $ThH_{10}$ [131], $CaH_6$ [16], $YH_9$ [101], and $LaH_{10}$ [101] closely match the experimental data [18,20,102–106], demonstrating excellent agreement between theory and experiment. It is worth noting that in 2023, two globally sensational room-temperature superconductors were reported: LK-99 [107] and the Lu-N-H system[108]. Despite numerous attempts by research teams to replicate these findings, none was successful [43,48,109–114].

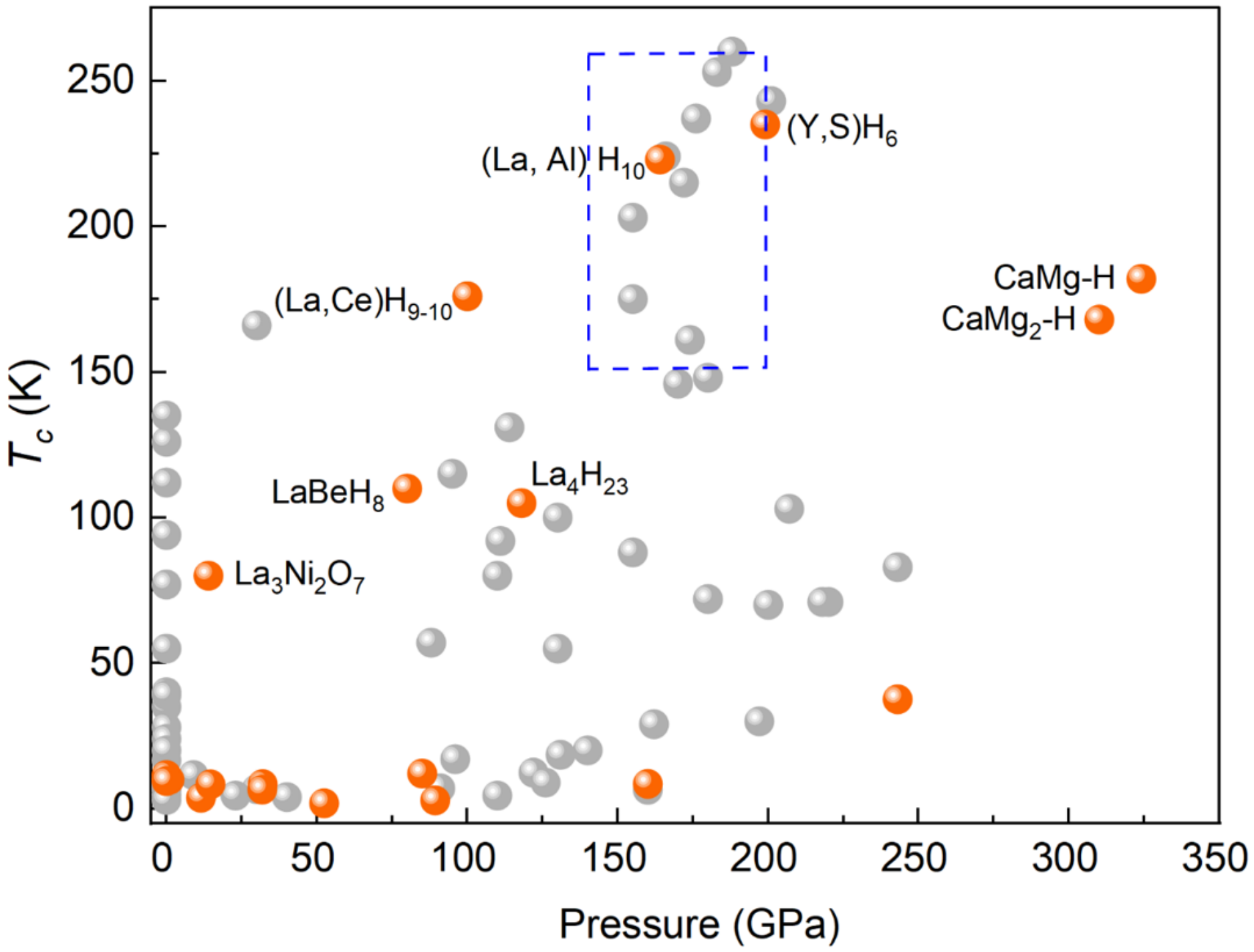

Fig. 6 $T_c$ and corresponding pressure of synthesized superconductors. The orange points represent the newly synthesized systems in 2023-2024[115–128]. The gray points represent synthesized superconductors before 2023.

At the end of 2023, Sun et al. observed that the layered perovskite nickelate $La_3Ni_2O_7$ [103], with the *Fmmm* space group, exhibited a high $T_c$ of up to 80 K at 14 GPa. Zhu et al. recently discovered superconductivity in another Ni-based material, the trilayer $La_4Ni_3O_{10-\delta}$ single crystal, with a $T_c$ of 30 K under 69 GPa[129]. These discoveries reignited interest in exploring the fundamental mechanisms of unconventional superconductors [130–132].

Song et al. synthesized $LaBeH_8$ in a diamond anvil cell at pressures between 110-130 GPa, measuring a $T_c$ of 110 K at 80 GPa[115]. This system features a La-Be framework and a novel $BeH_8$ unit stabilized by chemical compression from the La-Be scaffold. Inspired by $LaBeH_8$[115], researchers predicted many alternative doped structures, including $ThBeH_8$[58] ($T_c$ = 113 K at 69 GPa), $CeBeH_8$[58] ($T_c$ = 76 K at 28 GPa), and $YBeH_8$[59] ($T_c$ = 201 K at 200 GPa). New quaternary superhydrides based on the supercell structure of $LaBeH_8$, such as $ThLa_3Be_4H_{32}$ ($T_c$ = 134 K at 10 GPa), $LaAc_3Be_4H_{32}$ ($T_c$ = 135 K at 20 GPa), and $AcLa_3Be_4H_{32}$ ($T_c$ = 153 K at 20 GPa), were also proposed[40].

In the domain of high-pressure superconductivity, $LaH_{10}$, synthesized in 2019 at 188 GPa, continues to hold the record with a high $T_c$ of 260 K[106]. However, the requirement for extremely high stabilizing pressure poses significant experimental challenges. Scientists explored various methods, including element doping, to reduce the necessary pressure. For example, $(La,Al)H_{10}$ demonstrated a $T_c$ of 223 K at 164 GPa [119], while $(La,Ce)H_{10}$ exhibited a $T_c$ of 176 K at 100 GPa[116]. Prior theoretical investigations into LaH10's superconducting properties suggested a decrease in computed $T_c$ values with increasing pressure[133–135], contrary to experimental findings. This inconsistency is likely due to earlier calculations overlooking minor stoichiometric defects ($LaH_{9.6}$) in the actual structure, leading to quantum diffusion effects within the rigid lanthanum lattice and resulting in a positive pressure dependence of $T_c$[136].

Apart from hydrides and $La_3Ni_2O_7$, most superconductors synthesized in the past two years exhibited $T_c$ values mostly below 10 K. Examples include transition metal dichalcogenides such as $Re_{0.8}V_{0.2}Se_2$ [124] ($T_c$ = 3 K at 89 GPa), $V_{0.7}Re_{0.3}Se_2$ [125] ($T_c$ = 4 K at 11.5 GPa), and PdSSe[123] ($T_c$ = 12 K at 85 GPa); elemental superconductors Mo[120] ($T_c$ = 9 K at 160 GPa) and Sc [121] ($T_c$ = 38 K at 243 GPa), with Sc having the highest $T_c$ among known elemental superconductors. Additionally, a few topological insulator superconductors were discovered, including $GeBi_2Te_4$ [126] ($T_c$ = 8 K at 15 GPa), KHgAs[137] ($T_c$ = 7 K at 32 GPa), and $SnPSe_3$[138] ($T_c$ = 9 K at 32 GPa).

Discrepancies can sometimes exist between theoretically predicted $T_c$ and experimentally measured $T_c$ due to differences between theoretical models and actual experimental conditions. For instance, the predicted $T_c$ for $LiMoN_2$[36] could not be measured due to defects in the experimental structure[97]. The $T_c$ of (La,Ce)$H_{9-10}$ is difficult to accurately calculate because the computation cannot account for the configurational entropy introduced by disorder in the system[116]. The prediction for the $YH_9$ system was inaccurate due to a previously undiscovered experimental phase[54]. In many cases, the approach used in $T_c$ calculation, such as the McMillan equation [10], Allen-Dynes equation [11], and Eliashberg theory [14], can yield significant variations [139]. While the most accurate $T_c$ values should be used for comparison with experimental data, it is common to trade off accuracy for efficiency when dealing with a large number of systems. Furthermore, due to the high computational cost, the effects of anharmonicity and quantum nuclear effects are often neglected in calculations. However, these factors can also have a significant impact on $T_c$. Anharmonic corrections alter phonon frequencies, affecting both the stability of the system and the predicted $T_c$, while quantum nuclear effects primarily modify the potential energy surface, potentially allowing the system to remain dynamically stable at lower pressures, thereby influencing the phase diagram. For instance, in $CaH_6$, accurate $T_c$ predictions that match experimental values are only achieved when quantum nuclear and anharmonic effects are considered using the SSCHA method. [140] Similarly, while the $Fm$-$3m$ phase of $LaH_{10}$ is experimentally stable at 130 GPa, it is predicted to be unstable at 230 GPa under harmonic approximation, with anharmonic effects stabilizing it at lower pressures[135]. However, in some cases, such as $LaBH_8$, anharmonic effects can destabilize the system[141]. Thus, when comparing theoretical $T_c$ values with experimental data, it is important to examine the methods by which $T_c$ was determined and the detailed convergence of calculations.

As previously mentioned, while many ambient-pressure superconductors are computationally predicted to have high $T_c$, their experimental realization remains rare. This may be due to the fact that accurately predicting synthesizability is still challenging. Most crystal structure predictions (CSP) are performed at 0 K, where high-temperature effects are neglected, yet these effects can be critical in synthesis experiments [142]. Moreover, even if a material is thermodynamically stable, it can still be difficult to access in experiments if there is no suitable kinetic pathway. The computational design of reaction pathways to enhance the kinetic accessibility of predicted compounds has been proven to be both possible and critical [143] Future computational predictions should be closely integrated with experimental synthesis to form a feedback loop that enhances the synthesizability of predicted compounds. For instance, using computational design, Dolui et al. proposed a detour in the synthesis route to the high-$T_c$ superconductor $Mg_2IrH_6$ via $Mg_2IrH_x$ [61]. Recently, Hansen et al. successfully synthesized $Mg_2IrH_5$, which may serve as a starting point to access the novel $Mg_2IrH_6$ phase[144].

## IV. Conclusions

The search for new conventional superconductors emerged as a prominent field in computational materials science. The development of advanced computational tools led to numerous predictions of new superconducting compounds in recent years, underscoring the pivotal role of computational methods in identifying potential superconductors. Currently, the focus of computational predictions shifts towards ambient pressure conditions. Among the various candidates, hydrides and boron-carbides emerged as top choices due to their promising layered and cage-like structural prototypes[26]. The structural motifs and chemical compositions of these compounds play a crucial role in determining their superconducting properties. A critical finding is that $T_c$ is extremely sensitive to the involved chemical elements, their compositions, and structural motifs. Isoelectronic substitutions can significantly alter $T_c$, highlighting the complex interplay between material composition and superconducting properties. Therefore, a data-driven approach, complemented by machine learning, remains essential to explore a broader range of chemical compositions and structures to identify high-$T_c$ superconductors. Despite the progress in computational predictions, the experimental realization of these predicted superconductors is still lagging behind. Bridging this gap requires designing more effective synthesis routes to translate computational predictions into experimentally verified superconductors. Addressing this challenge is crucial and urgent for advancing the field and achieving practical applications of new superconducting materials.

## Acknowledgements

We are grateful to W. E. Pickett for valuable suggestions on the manuscript. The work at Xiamen University was supported by the Natural Science Foundation of Xiamen (Grant No. 3502Z202371007) and the Fundamental Research Funds for the Central Universities (Grant No. 20720230014). The work at Jimei University was supported by the Natural Science Foundation of Xiamen City (Grant No. 3502Z202372015) and the Research Foundation of Jimei University (Grant No. ZQ2023013). V.A. was supported by the U.S. Department of Energy, Office of Basic Energy Sciences, Division of Materials Sciences and Engineering. Ames National Laboratory is operated for the U.S. Department of Energy by Iowa State University under Contract No. DE-AC02-07CH11358. K.-M.H. acknowledges support from National Science Foundation Award No. DMR2132666.